\documentclass[submission,copyright,creativecommons]{eptcs}
\providecommand{\event}{FROM 2026} 

\usepackage{amsmath,amsthm,amssymb,graphicx}
\usepackage{listings}
\usepackage{xcolor}
\usepackage{tabularx}
\usepackage{multirow}
\usepackage{caption}
\usepackage[english]{babel}
\usepackage{makecell}
\usepackage{tikz}
\usepackage{fancyvrb}
\usetikzlibrary{arrows.meta, positioning, fit, backgrounds, shapes.geometric}
\usepackage{paralist}
\usepackage{float}
\RequirePackage{hyphenat}
\usepackage{algorithm, algpseudocode}
\usepackage[inline]{enumitem}
\usepackage{url}
\usepackage{comment}
\usepackage[
]{hyperref}

\usepackage{titlesec}

\titlespacing*{\paragraph}
  {0pt}    
  {0.6em}  
  {0.6em}  

\usepackage{iftex}

\ifpdf
  \usepackage[T1]{fontenc}        
\else
  \usepackage{breakurl}           
\fi

\definecolor{tlaKeyword}{RGB}{0,76,153}
\definecolor{tlaProof}{RGB}{124,45,141}
\definecolor{tlaConstant}{RGB}{166,68,0}
\definecolor{tlaOperator}{RGB}{0,112,95}
\definecolor{tlaComment}{RGB}{80,110,80}
\definecolor{tlaString}{RGB}{150,40,40}
\definecolor{tlaNumber}{RGB}{110,110,110}
\definecolor{tlaFrame}{RGB}{205,205,205}
\definecolor{codegreen}{rgb}{0,0.6,0}
\definecolor{codegray}{rgb}{0.5,0.5,0.5}
\definecolor{codepurple}{rgb}{0.58,0,0.82}
\definecolor{backcolour}{rgb}{0.95,0.95,0.92}

\lstdefinelanguage{TLA+}{
  sensitive=true,
  alsoletter={\\},
  morekeywords=[1]{
    ASSUME,ASSUMPTION,AXIOM,CASE,CHOOSE,CONSTANT,CONSTANTS,
    DOMAIN,ELSE,ENABLED,EXCEPT,EXTENDS,IF,IN,INSTANCE,LAMBDA,
    LET,LOCAL,MODULE,OTHER,RECURSIVE,SF_,SUBSET,THEN,THEOREM,
    UNCHANGED,UNION,VARIABLE,VARIABLES,WF_,WITH
  },
  morekeywords=[2]{
    ACTION,BY,COROLLARY,DEF,DEFINE,DEFS,HAVE,HIDE,LEMMA,NEW,
    OBVIOUS,OMITTED,ONLY,PICK,PROOF,PROPOSITION,PROVE,QED,
    STATE,SUFFICES,TAKE,TEMPORAL,USE,WITNESS
  },
  morekeywords=[3]{
    TRUE,FALSE,BOOLEAN,STRING
  },
  morekeywords=[4]{
    \A,\E,\AA,\EE,
    \approx,\asymp,\bigcirc,\bullet,\cap,\cdot,\circ,\cong,
    \cup,\div,\doteq,\equiv,\geq,\gg,\in,\intersect,\land,
    \leq,\ll,\lnot,\lor,\neg,\notin,\o,\odot,\ominus,\oplus,
    \oslash,\otimes,\prec,\preceq,\propto,\sim,\simeq,
    \sqcap,\sqcup,\sqsubset,\sqsubseteq,\sqsupset,
    \sqsupseteq,\star,\subset,\subseteq,\succ,\succeq,
    \supset,\supseteq,\times,\union,\uplus,\wr,\X
  },
  morecomment=[l]{\*},
  morecomment=[n]{(*}{*)},
  morestring=[b]"
}

\lstdefinestyle{tla}{
  language={TLA+},
  backgroundcolor=\color{backcolour},   
  upquote=true,
  basicstyle=\fontencoding{T1}\footnotesize\ttfamily,
  keywordstyle=[1]\color{tlaKeyword}\bfseries,
  keywordstyle=[2]\color{tlaProof}\bfseries,
  keywordstyle=[3]\color{tlaConstant}\bfseries,
  keywordstyle=[4]\color{tlaOperator},
  commentstyle=\color{tlaComment}\itshape,
  stringstyle=\color{tlaString},
  numbers=left,
  numberstyle=\tiny\color{tlaNumber},
  numbersep=8pt,
  frame=single,
  rulecolor=\color{tlaFrame},
  columns=fullflexible,
  keepspaces=true,
  captionpos=b,
  showstringspaces=false,
  upquote=true,
  tabsize=2,
  breaklines=true,
  literate=*
    {WF_}{{{\color{tlaKeyword}\bfseries WF\_}}}3
    {SF_}{{{\color{tlaKeyword}\bfseries SF\_}}}3
}
\title{Monitoring and Verification of Multitenant Kubernetes Clusters using TLA\textsuperscript{+}  Trace Checking}
\author{
Ioana Sila\c{s} \qquad\qquad Adrian Cr\u{a}ciun
\institute{Faculty of Informatics\\
West University Timi\c{s}oara\\
Romania}
\email{\quad ioana.silas01@e-uvt.ro \quad\qquad adrian.craciun@e-uvt.ro}
}
\def\titlerunning{Monitoring and Verification of Multitenant Kubernetes
Clusters using TLA+ Trace Checking}
\def\authorrunning{I. Sila\c{s} \& A. Cr\u{a}ciun}
\begin{document}
\maketitle

\begin{abstract}
In distributed systems, model checking is usually used at design time for specifying an abstract model of the system and then exhaustively checking all possible behaviors. TLA\textsuperscript{+} is commonly used in this way as a specification language, together with the TLC model checker. 

In this paper, we present a monitoring tool that, at its core, utilizes TLA\textsuperscript{+} specifications in a different way. The tool utilizes a TLA\textsuperscript{+} trace-checking specification to detect violations in behavior inferred from Kubernetes audit logs. Our primary use case focuses on multitenancy violations; however, the pipeline is not limited to that setting. Specifically, it demonstrates how formal reasoning can be incorporated into live Kubernetes environments to improve monitoring and correctness checking.
\end{abstract}

\section{Introduction}

Kubernetes (K8s), see ~\cite{K8sDocs}, is a widely-used, open-source container orchestration platform. It automates the deployment, scaling, and management of containerized applications and has become the de facto standard for many cloud-native environments. The complexity of K8s introduces challenges in ensuring the correctness and security of live systems. Another layer of complexity is added by \emph{multitenancy}, which allows multiple teams or applications to share the same cluster to reduce infrastructure costs and improve resource utilization. This setup introduces potential security and stability concerns, since it can lead to interference between different tenants in the cluster, whether malicious or unintentional. To mitigate these risks, it is essential to enforce strict isolation mechanisms across namespaces and resources.

Formal methods have traditionally been used to model and verify distributed systems in order to provide strong guarantees of correctness and security. TLA\textsuperscript{+}, see~\cite{SpecifyingSystems}, a formal specification language, has been widely used in the industry by companies such as Intel, see~\cite{batsonleslie2002}, Amazon, see~\cite{Newcombe2015}, and others.

In this paper, we use  TLA\textsuperscript{+} to create a model of a multitenant cluster: we describe predicates, actions, invariants. Then, we extend the role of specifications that reflect the cluster states by using them to monitor live cluster behavior. In this way, we can confirm that the role-based access control (RBAC) implementation fits our abstracted cluster design, which otherwise would be difficult to validate.

To achieve this, we create a tool that integrates cluster monitoring with formalism. Rather than just performing exhaustive model checking (as seen in traditional TLA\textsuperscript{+} applications), our approach verifies that observed system behaviors align with a predefined specification. The goal is to shift the focus from abstract verification to concrete, runtime observation, in order to provide a solution that is both accessible to system administrators and more rigorous than standard monitoring tools. Such tools provide visibility into runtime events but lack formal correctness guarantees.  

\pagebreak

In order to achieve our goal, we take the following steps:

\begin{itemize}
    \item specify expected K8s multitenant cluster behavior using TLA\textsuperscript{+}, focusing on state transitions and invariants,
    \item extract and normalize execution traces from K8s logs,
    \item check the reconstructed trace against the TLA\textsuperscript{+} specification using a monitoring pipeline,
    \item evaluate the approach in multitenant scenarios to assess scalability and the pipeline's ability to detect violations. 
\end{itemize}

What we want to achieve is a combination of formal modeling with practical system observability in order to link specification-level correctness and real-world K8s behavior. The goal is to implement continuous validation of multitenant deployed clusters while preserving the operational semantics of K8s.

Our contributions are: TLA\textsuperscript{+} specifications for multitenant K8s, a monitoring pipeline that can be deployed in clusters. In addition, we have written custom TLA\textsuperscript{+} operators that interact with a messaging system and its persistence layer. We have also deployed a \texttt{kubeadm} K8s cluster locally on virtual machines and used it to validate the operational flow of the pipeline.

The paper is organized as follows: Section~\ref{formal-model} presents an overview on K8s multitenancy and the TLA\textsuperscript{+} specifications at the core of the pipeline. Section~\ref{PipelineArchitecture} describes the architecture and design of the monitoring pipeline. Section~\ref{ExperimentalResults} details the experimental setup, methodology, and results. Section~\ref{RelatedWork} reviews related work in trace checking and K8s multitenancy. Finally, Section~\ref{ConclusionFW} discusses the findings of this paper and directions for future improvements. The implementation of the monitoring pipeline discussed throughout this paper and the experimental results are available in our GitHub repository\footnote{\url{https://github.com/zwx13/k8s-runtime-audit/tree/main}}.

\section{Modeling K8s Multitenancy in TLA\textsuperscript{+}}
\label{formal-model}

We introduce TLA\textsuperscript{+} and explain the particular way we used it here -- \emph{trace checking}. We then explain how we formalized a particular model of K8 multitenancy, in order to monitor multitenant K8 clusters. 

\subsection{Formal Instrument: TLA\textsuperscript{+}}
\label{TLAOverview}

TLA\textsuperscript{+} is a formal language based on set theory, which focuses on model checking (see \cite{SpecifyingSystems}). It is based on a state machine model: we describe an initial state and specify the possible transitions through which the system can mode from one state to the next.

The usual usage of the language follows this trajectory: a specification that describes the essential behavior of an abstracted away algorithm or system is written, then it is model-checked. If model-checking reports a violation, the counterexample can be used to improve the specification or modify the design before the system implementation. Another case is when the system is already implemented, but writing a specification leads to uncovering hidden assumptions or design flaws that were not obvious from code or testing. In both cases, however, the implementation is not linked directly to the specification, since a highly abstractized version is created and model-checked. 

This limitation is discussed in \cite{ValidateTraces}, where the authors mention that while TLA\textsuperscript{+} and similar formal languages provide mechanisms to check an abstracted version of the system, they do not provide means to check the correctness of the actual implementation. As a result, another use of TLA\textsuperscript{+}, and the one we adopt here, is to take the highly abstracted specification and use it to verify traces produced by the implemented system. 
By \emph{trace checking} we reuse the initial specification of the system (K8 multitenancy in our case) for the purpose of monitoring deployed multitenant clusters. 

Note that trace checking does not prove that the implementation or the specification is correct. It solely checks whether recorded executions are compatible with the specification. This depends on what the trace or log specifies, the relation between the specification and implementation. Logging the relevant events at the appropriate level of atomicity can become costly if the implementation is very complex (as seen in \cite{ExtremeModeling2020}). As a result, we should look at trace checking not as a proof of implementation correctness, but rather as a way of reusing TLA\textsuperscript{+} specifications at the implementation level and identifying differences between the model and actual execution.

The TLA\textsuperscript{+} use in the work presented here can be summarized as follows: \begin{inparaenum}[(1)]
\item \emph{The base specification} represents the abstract model of the multitenant cluster and the intended multitenant policy. The model was constructed based on the K8 documentation covering multitenancy and Subsection~\ref{base-spec} provides an overview of this model. The development of the model included the use of the TLC, the model checker of TLA\textsuperscript{+} to ensure consistency, which is the standard way TLA\textsuperscript{+} is used. However, the consistency of the model is not the focus of the work presented in this paper and we skip the details.  
\item \emph{The trace specification} reuses the base specification's predicates and derived sets to classify reconstructed states as allowed or disallowed. 

The base specification provides the policy semantics, while the trace specification provides the observational replay semantics. This is not refinement or conformance checking, but rather TLA\textsuperscript{+}-based audit trace checking or runtime policy monitoring. We replay observed events, detect bad states by using predicates, and the replay continues so violations can be accumulated instead of stopping at the first non-conforming transition. This is explained in Subsection~\ref{TraceSpec}. 
\end{inparaenum}

\subsection{Proposed Multitenancy Model: The Base Specification}
\label{base-spec}

There is no single standard model when it comes to K8s multitenancy. We specify \emph{a representative soft tenancy model} which is built on \texttt{Namespace} isolation between tenants and role-based access control.  

K8s resources can be either cluster-wide or scoped to a \texttt{Namespace}. \texttt{Namespaces} aid in dividing cluster resources between multiple teams. In soft multitenancy they provide a grouping mechanism for workloads. For example, a team can use one or more \texttt{Namespaces} to deploy their applications, and labels on the \texttt{Namespaces} can identify ownership and tenant information. This is useful in creating the \emph{tenant} abstraction, since K8s does not have a native \texttt{Tenant} object. \texttt{Namespaces} are created by cluster administrators since they are global objects.

\texttt{Roles} define permissions at the \texttt{Namespace}-level, while \texttt{ClusterRoles} define permissions at the cluster level, or reusable permissions that can be used inside a \texttt{Namespace}. In either case, the permissions become effective only when they are bound through \texttt{RoleBindings} or \texttt{ClusterRoleBindings}. The former have \texttt{Namespace}-wide effect, while the latter's scope is the whole cluster.

\begin{lstlisting}[float = h, caption={The base specification - multitenancy concepts and constraints.},label={typeok-base}]
TypeOK ==   
        /\ nsTenantMap \in [Namespaces -> (Tenants \cup {NoTenant})]
        /\ DOMAIN roleBindings \in SUBSET (Namespaces \X RBNames)
        /\ \A key \in DOMAIN roleBindings: roleBindings[key] \in (Groups \X ClusterRoleNames)
        /\ DOMAIN clusterRoleBindings \in SUBSET CRBNames
        /\ \A key \in DOMAIN clusterRoleBindings: 
                    clusterRoleBindings[key] \in (Groups \X ClusterRoleNames) 
        /\ DefaultClusterRoleNames \in SUBSET DOMAIN clusterRoles
        /\ DOMAIN clusterRoles \in SUBSET ClusterRoleNames
        /\ \A key \in DOMAIN clusterRoles: clusterRoles[key] \in Permissions
        /\ DOMAIN accessAttempts \in SUBSET (Namespaces \X Groups \X Permissions)
        /\ \A key \in DOMAIN accessAttempts: 
        accessAttempts[key] \in [respectsNSTMapAtReqTime: BOOLEAN, matchingRBorCBR: BOOLEAN]
\end{lstlisting}

The \texttt{TypeOK} state predicate defines these RBAC \textbf{concepts} in our model, see Listing~\ref{typeok-base}.

Permissions can be granted at the group level with either \texttt{ClusterRoleBindings} or \linebreak \texttt{RoleBindings}. These bindings can bind only to \texttt{ClusterRoles}, not to \texttt{Roles}. We chose to follow a strict version of the principle of least privilege (see \cite{K8sRBACPractices}) and did not include the latter in our model at all. Allowing arbitrary \texttt{Roles} in each \texttt{Namespace} would increase the number of possible permission configurations without adding much value for the representative model. The default K8s \texttt{ClusterRoles} already cover the common access levels needed in this model, while cluster administrators can still create custom \texttt{ClusterRoles} if more specific permissions are required.

\begin{lstlisting}[float=h, caption={Possible actions - transitions of the system.}, label={base-next}]
Next == \E admingroup \in Groups, rbName \in RBNames, crbName \in CRBNames, targetgroup \in Groups, ns \in Namespaces, t \in Tenants, p \in Permissions, cr \in (ClusterRoleNames):
            \/ CreateNamespace(admingroup, ns, t)
            \/ DeleteNamespace(admingroup, ns)
            \/ CreateClusterRole(admingroup, cr, p)
            \/ UpdateClusterRole(admingroup, cr, p)
            \/ DeleteClusterRole(admingroup, cr)
            \/ GrantNSAccess(admingroup, ns, rbName, targetgroup, cr)
            \/ RevokeNSAccess(admingroup, ns, rbName, targetgroup, cr)
            \/ GrantClusterAccess(admingroup, crbName, targetgroup, cr)
            \/ RevokeClusterAccess(admingroup, crbName, targetgroup, cr)
            \/ AttemptAccess(ns, targetgroup, p)
\end{lstlisting}

\textbf{The actions} (transitions of the multitenant cluster) are described by the \texttt{Next} predicate, as illustrated in Listing~\ref{base-next}. These actions describe creation and deletion of namespaces, creation, deletion, updates to cluster roles, granting and revocation of namespace or cluster access, as well as attempted access to resources of particular namespaces.

The cluster administrators are the only ones who can create custom \texttt{ClusterRoles}. Besides these, the cluster administrators can use one of the four default \texttt{ClusterRoles}: \texttt{view}, \texttt{edit}, \texttt{admin}, and \texttt{cluster-admin}. When these are bound inside a \texttt{Namespace} with a \texttt{RoleBinding}, even if the permissions are wide, such as those granted by assigning the \texttt{ClusterRole} of \texttt{cluster-admin}, they are \texttt{Namespace}-scoped. When the \texttt{ClusterRoles} are granted through \texttt{ClusterRoleBindings}, they are cluster-scoped.

As for the hierarchy inside \texttt{Namespaces}, in each of them there can be a group that is assigned the \texttt{ClusterRole} that grants admin powers. Such a group has permissions to create \texttt{RoleBindings} only, but not \texttt{ClusterRoleBindings}, to give rights to other groups that operate in the same \texttt{Namespace}. However, abiding by the principle of least privilege, no group can grant more powers than they have. As a result, \texttt{Namespace} administrators can only create \texttt{RoleBindings} that bind to \texttt{view} or \texttt{edit}. Cluster administrators can grant whatever \texttt{ClusterRole} they choose to in any \texttt{Namespace}. See Listing~\ref{base-grantAccess}.

\begin{lstlisting}[float = h, caption={Granting namespace access.}, label={base-grantAccess}]
GrantNSAccess(admingroup, targetNS, rbName, targetG, cr) ==
            /\ cr \in DOMAIN clusterRoles
            /\     LET clusterPerm == clusterRoles[cr]
                    IN \/   /\ IsClusterAdmin(admingroup)
                            /\ clusterPerm \in {"none", "read", "write", "admin-powers"}

                       \/   /\ IsNSAdmin(admingroup, targetNS)
                            /\ clusterPerm \in {"none", "read", "write"}
            /\ SameTenant(targetNS, targetG)
            /\ roleBindings' = <<targetNS, rbName>> :> <<targetG, cr>> @@ roleBindings
            /\ UNCHANGED << nsTenantMap, clusterRoleBindings, clusterRoles, accessAttempts >>
\end{lstlisting}

The \texttt{AttemptAccess} action, see Listing~\ref{base-attemptaccess},  models the event where a group attempts to access resources in a particular \texttt{Namespace}. The access attempts are monitored and they also note whether they were allowed or not, based on the actor group, the relation between this group and the \texttt{Namespace}, and the existence of a matching \texttt{(Cluster)RoleBinding}.

The details concerning the rest of the actions are skipped for the purpose of this presentation. The code is available in our repository. 

\begin{lstlisting}[float = h, caption={Modeling access attempt.}, label={base-attemptaccess}]
AttemptAccess(ns, g, p) ==
  /\ nsTenantMap[ns] # NoTenant
  /\ accessAttempts' = 
        IF <<ns, g, p>> \in DOMAIN accessAttempts THEN
          [accessAttempts EXCEPT 
           ![<<ns, g, p>>].respectsNSTMapAtReqTime = 
                    (SameTenant(ns, g) \/ g \in AdminGroups),
           ![<<ns, g, p>>].matchingRBorCBR = 
                    (MatchRoleBinding(ns, g, p) \/ MatchCRBinding(g, p))]
        ELSE <<ns, g, p>>  :> 
          [respectsNSTMapAtReqTime |-> (SameTenant(ns, g) \/ g \in AdminGroups),
           matchingRBorCBR |-> (MatchRoleBinding(ns, g, p) \/ MatchCRBinding(g, p))
          ] @@ accessAttempts
  /\ UNCHANGED << nsTenantMap, roleBindings, clusterRoleBindings, clusterRoles >>
\end{lstlisting}

\textbf{The invariants} described in the base specification focus on common multitenancy violations: cross-tenant bindings, dangling bindings, cross-tenant successful access attempts, \texttt{RoleBindings} to \linebreak \texttt{cluster-admin}, no \texttt{ClusterRoleBindings} for tenant groups, see Listing~\ref{base-invariants}. Listing~\ref{base-noCrossTSuccess} illustrates cross tenant access, the rest are available in the repository. 
\begin{lstlisting}[float=h, caption={Invariants for the base specification.}, label={base-invariants}]
Inv ==  /\ TypeOK 
        /\ BindingsRespectMT 
        /\ NoCrossTenantSuccess 
        /\ NoDanglingBindings 
        /\ NoClusterAdminRB 
        /\ NoTenantCRB
\end{lstlisting}

\begin{lstlisting}[float = h, caption={Example invariant: groups cannot access ourside their tenant, unless they are cluster admins.}, label={base-noCrossTSuccess}]
NoCrossTenantSuccess ==  \A a \in DOMAIN accessAttempts :
                            LET group == a[2]
                            IN accessAttempts[a].matchingRBorCBR = TRUE => 
                                accessAttempts[a].respectsNSTMapAtReqTime = TRUE
\end{lstlisting}

But our objective is to deal with deployed multitenant clusters. Any violation of the invariants from the base specification would lead to a halt to the system. Enters...  

\subsection{The Trace Specification}
\label{TraceSpec}
The trace specification differs from the base specification in both purpose and input. As we have seen, the base specification models the desired state and allowed transitions of the system. In contrast, the trace specification operates on K8s audit logs directly. Its role is not to generate all possible system behaviors, but to replay observed events and check whether they violate the multitenancy rules we have defined.

Model checking is a finite process, so we need to process logs in batches we can reasonably deal with. Each log entry is represented in JSON, but the trace specification needs to access fields such as the request's verb, resource, \texttt{Namespace} where it took place, user who created the request. In order to achieve this, we convert the logs to a type that TLA\textsuperscript{+} understands. Moreover, since the trace specification operates on batches of logs, we preserve the intermediary state between batches. This allows the next batch to pick up from the state reached by the previous specification. In addition, detected violations are written separately as alerts so that cluster administrators can investigate them.

To achieve this, we created custom operators that bridge the TLA\textsuperscript{+} specification and outer programs (for more details, see Subsection~\ref{NATSOperators}). The TLA\textsuperscript{+} operators retrieve the relevant system logs in a deterministic manner. Since model checking is not a single linear execution, a state or transition can be revisited from different paths. As a result, TLC may evaluate operators multiple times while exploring the state space. The main idea around the operators we implemented is that the log retrieval mechanism is stable and they are idempotent: operators always fetch the same batch of logs for a given state. 

The initial state initializes the \texttt{idx} variable to 1 (first element in a batch). We then check if there is a saved state to be picked-up from, and if so, we use it as a checkpoint. If there is not, we initialize \texttt{Init} to match the base specification, see Listing~\ref{trace-init}. 
\begin{lstlisting}[float = h, caption = {Initial state of the trace specification.}, label = {trace-init}]
Init == 
    /\ idx = 1
   ...
    IF \/ IsEmpty \/ HasEmptyMappings
    THEN   (* InitBase*)
      /\ nsTenantMap = [ ns \in Namespaces |-> NoTenant ]
      /\ roleBindings = [nsrb \in {} |-> {}]
      /\ clusterRoleBindings = "cluster-admin":> <<"kubeadm:cluster-admins","cluster-admin">>
      /\ clusterRoles = DefaultClusterRolePermMap
    ELSE (* InitFromCheckoint*)
      /\ nsTenantMap = IF HasEmptyNSTenantMap THEN [ ns \in Namespaces |-> NoTenant ]
                                              ELSE AllocIn.nsTenant
      /\ clusterRoles = AllocIn.clusterRoles
      /\ roleBindings = AllocIn.roleBindings
      /\ clusterRoleBindings = AllocIn.clusterRoleBindings
\end{lstlisting}

In the \texttt{Next} state, we update variables directly. We do not call the matching actions from the base specification because the trace specification should allow actions that the base does not. Specifically, audit logs can contain violations of the desired cluster state reflected by the base specification. These events are real cluster events, so they should be accepted by the trace specification. For the same reason, the trace specification does not use invariants. Instead, it contains alert actions that add data about the violating event in a set, after it is encountered. These alerting operators are based on the base specification invariants. The trace specification records policy violations as alerts while continuing to process the audit logs in the batch.

\begin{lstlisting}[float = h, caption = {Alert processing example.}, label={trace-alert}]
CrossTenantSuccessSet == {a \in DOMAIN accessAttempts :
        /\ accessAttempts[a].matchingRBorCBR
        /\ ~(accessAttempts[a].respectsNSTMapAtReqTime)}

AlertIfCrossTenantAction ==
    LET crossTenantAction == Model!CrossTenantSuccessSet' \ Model!CrossTenantSuccessSet IN
        IF crossTenantAction = {} THEN 
            /\ TRUE
            /\ UNCHANGED << crossTenantAlerts >>
        ELSE 
            /\ crossTenantAlerts' = crossTenantAlerts \cup { << LogEvents[idx]["auditID"], LogEvents[idx]["tlaType"] >> }
            /\ PrintT("!!! Cross Tenant Access Identified !!!")
\end{lstlisting}

The basis for the alerting mechanism is the difference between the violation set before and after processing an event (e.g. \texttt{CrossTenantSuccessSet}). If the difference is empty, the current event did not introduce a new violation. The trace specification adds an alert with details from the relevant audit event if the difference is not empty. The purpose of this mechanism is to avoid reporting the same previously known violation repeatedly, so that the alert can correspond to the event that actually violates the rules we established in the base specification. This is illustrated in Listing~\ref{trace-alert}.

As for the allowed behavior, the \texttt{AlertIfBadState} action and \texttt{Next} action are combined into one. Processing a log entry can affect both the variables that represent the modeled cluster state and those that store detected alerts. The two actions are complementary: \texttt{Next} updates the cluster-state variables according to the information from the currently processed audit event, while \texttt{AlertIfBadState} evaluates the resulting state and updates just the alert sets: it does not modify the cluster-state variables. We combine the two actions using conjunction since they operate on disjoint groups of variables.

\vspace{0.2cm}
{
\begin{minipage}{0.47\textwidth}
\begin{lstlisting}
AlertIfBadState == 
  /\ AlertIfCrossTenantAction 
  /\ AlertIfDanglingRoleBindings 
  /\ AlertIfDanglingClusterRoleBindings 
  /\ AlertIfClusterRoleBindingForTenant 
  /\ AlertIfRoleBindingToClusterAdmin
\end{lstlisting}

\end{minipage}
\hfill
\begin{minipage}{0.47\textwidth}
\begin{lstlisting}
NextPrintSerialize == 
    PrintInitOnce \/ (Next /\ AlertIfBadState) \/ SerializeAtEnd

TraceBehavior == 
    Init /\ [][NextPrintSerialize]_vars
\end{lstlisting}
\end{minipage}
 \captionof{lstlisting}{Alerts, transitions in the trace specification.}
    \label{alertbadstate}
}

\section{Pipeline Architectural Details}
\label{PipelineArchitecture}

The monitoring pipeline is designed as a K8s-native tool, so administrators can deploy it directly inside their clusters. Deployment is managed through Helm (see \cite{HelmDocs}), which provides a package-like mechanism for installing and configuring K8s resources. The framework is organized as a pipeline of independent components which include a Python webhook, stream creating scripts, filtering scripts, and a Java-based TLC runner. These components are decoupled, so each stage can be updated or replaced independently, without disrupting the whole monitoring flow.

\begin{figure}[ht]
\centering
\resizebox{0.85\textwidth}{!}{%
\begin{tikzpicture}[
    node distance=0.7cm and 1.2cm,
    block/.style={
        draw,
        rounded corners,
        align=center,
        minimum width=2.4cm,
        minimum height=0.75cm,
        font=\scriptsize\sffamily,
        fill=white,
        inner sep=2pt
    },
    stream/.style={
        draw,
        circle,
        minimum size=1.25cm,
        align=center,
        font=\tiny\sffamily,
        fill=gray!5,
        inner sep=1pt
    },
    group/.style={
        draw,
        dashed,
        inner sep=0.35cm,
        rounded corners,
        fill=blue!5,
        fill opacity=0.15
    },
    k8sgroup/.style={
        draw,
        dotted,
        inner sep=0.3cm,
        rounded corners,
        fill=gray!5,
        fill opacity=0.2
    },
    ->, >=stealth, thick
]

\node[block] (api) {K8s API Server};
\node[block, right=of api] (audit) {Audit Log\\Mechanism};

\node[block, below=2.2cm of audit] (webhook) {Webhook};

\node[stream, below=0.7cm of webhook] (raw_stream) {Audit\\Stream};
\node[stream, left=of raw_stream] (multi_stream) {MT\\Stream};

\node[block, left=of multi_stream] (java) {Trace\\Runner};
\node[block, below=1cm of java] (tlc) {TLC \\(TLA+ Model-Checker)};

\node[stream, left=1.1cm of tlc] (alerts) {Alerts\\Stream};
\node[stream, right=1.1cm of tlc] (kv) {Storage\\Stream};

\draw (api) -- (audit);
\draw (audit) -- (webhook) node[midway, right, font=\tiny] {HTTPS POST};
\draw (webhook) -- (raw_stream);
\draw (raw_stream) -- (multi_stream) node[midway, above, font=\tiny\bfseries] {Filter};
\draw (multi_stream) -- (java) node[midway, above, font=\tiny] {Ingest};
\draw (java) -- (tlc);
\draw (tlc) -- (alerts);
\draw (tlc) -- (kv);

\begin{scope}[on background layer]
    \node[k8sgroup, fit=(api) (audit)] (k8szone) {};
    \node[anchor=north west, font=\scriptsize\itshape, yshift=0.4cm] at (k8szone.north west) {Native K8s Components};

    \node[group, fit=(webhook) (raw_stream) (multi_stream) (java) (tlc) (alerts)  (kv)] (myzone) {};
    \node[anchor=north west, font=\scriptsize\bfseries] at (myzone.north west) {TLA\textsuperscript{+} Monitoring Pipeline};

    \node[draw, line width=0.8pt, inner sep=0.55cm, fit=(k8szone) (myzone)] (cluster) {};
    \node[anchor=north east, font=\normalsize\sffamily\bfseries, yshift=0.45cm] at (cluster.north east) {K8s Cluster};
\end{scope}

\end{tikzpicture}%
}
\vspace{-0.5em}
\caption{Architecture of the integrated monitoring pipeline within the K8s cluster.}
\label{fig:full_cluster_architecture}
\end{figure}
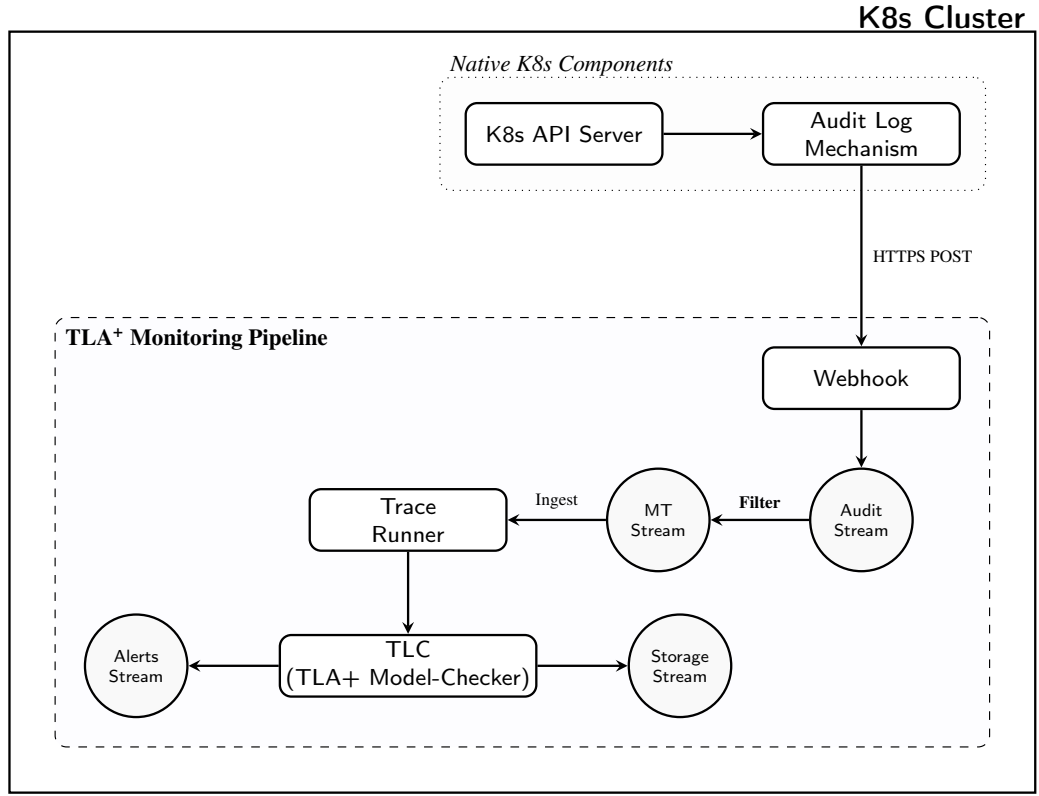

The flow of the pipeline is illustrated in Figure~\ref{fig:full_cluster_architecture} and goes as follows. The K8s API server continuously produces audit logs for cluster activity. By default, these logs are generated by the API server, but they are not automatically forwarded to an external processing component. For this reason, we configured a webhook endpoint where audit events are received.

After receiving the audit events, the webhook forwards them to an \texttt{Audit} stream in NATS Jetstream (see \cite{NatsJetstream}), which is used as a high-performance messaging system. From there, the logs are consumed and further filtered to reduce noise and keep only those events that are relevant to the multitenancy model. Then we have another stream that acts as storage for cleaned and classified logs. We call this stream the \texttt{Multitenancy} stream. Before logs are sent to this stream, we classify them to match the abstracted permissions in the TLA\textsuperscript{+} specifications. After this classification, the logs are prepared for analysis and consumption by the TLC runner as input for trace checking.

The pipeline depends on an ordering guaranty regarding the audit events. This assumption has two parts. First, K8s audit logs are treated as a chronological record of API-server activity (see \cite{K8sAudit}). Second, once these events are received by the webhook, they are published to a single NATS JetStream stream. JetStream streams store messages with sequence numbers, and TLC consumes messages from the \texttt{Multitenancy} stream in order. The trace checked by TLC corresponds to the order in which the relevant audit events are stored in the final stream.

Batches of audit logs are processed by sequential TLC invocations, with a one-second interval between runs; TLC operates continuously as part of the monitoring pipeline.

When the pipeline is first deployed, a Kubernetes job performs an initial catch-up with the current cluster state and stores the resulting state in NATS JetStream. Subsequent batches of audit logs are then processed using the TLA\textsuperscript{+} trace specification, starting from the previously stored state.

After each TLC run, two outcomes are possible. If a specification violation is detected, the relevant audit logs are published to the \texttt{Alerts} stream so that the violation can be inspected and appropriate clean-up actions can be performed. The final model state is also persisted. If no violation is detected, no alert is emitted, but the final state is still stored persistently in NATS JetStream. This stored state acts as a checkpoint for the next TLC invocation: it allows each new batch of logs to be checked relative to the state produced by the previous batch.

\subsection{Custom NATS Jetstream TLA\textsuperscript{+} Operators}
\label{NATSOperators}
As mentioned in Sunbsection~\ref{TraceSpec}, the TLA\textsuperscript{+} trace specification interacts with an external system to fetch batches of audit logs, save state between runs, and output alerts that can then be inspected by cluster administrators. This is achieved with the help of custom operators implemented in Java, that are then called inside the trace specification.
 
\paragraph{\texttt{NatsConsume}:} This is the log-fetching operator. The idea is that the fetching mechanism must remain stable: the operator must always fetch the same batch of logs for a given state, instead of advancing through the log stream during a revisiting of states. This is achieved with the help of a flag that is set to \texttt{True} once successful retrieval of the batch is finished. If the operator is called again during the same TLC process, it fetches the same batch instead of advancing and acking the next set of messages that might have arrived in the meanwhile.

\paragraph{\texttt{NatsLoadCachedState}:} This operator is used inside the initial state of the trace specification to retrieve the current state saved in the \texttt{Storage} stream in NATS JetStream. The pipeline could be deployed inside the cluster at any point: we created a mechanism that when the \texttt{Storage} stream is created, an entry with the current relevant cluster objects is added to it. Moreover, at the end of a TLC process, the state is saved in the same \texttt{Storage} stream, overwriting the previous entry. Its purpose is to ensure that the loaded cluster state from storage accurately reflects the environment. 

\paragraph{\texttt{NatsPutCachedState}:} This is the counterpart to the previous operator. When a TLC process finishes executing, the resulting state is saved inside the \texttt{Storage} stream, so it can then be loaded when the next process runs. Both operators are idempotent, similar to the \texttt{NatsConsume}: if the  \texttt{NatsLoadCachedState} operator's flag indicates that the state was already fetched once during the same process, it returns the same state, instead of trying to load it again. As for the \texttt{NatsPutCachedState} operator, if it gets called again during model-checking, it returns \texttt{True} instead of re-publishing.

\paragraph{\texttt{NatsPublishAlert}:}Finally, this operator publishes alerts to the \texttt{Alerts} stream. These alerts are based on the mechanism showcased in ~\ref{TraceSpec}. Similar to the \texttt{NatsPutCachedState} operator, after alerts have been published once during a TLC process, the operator returns \texttt{True} instead of publishing them again.

\section{Experimental Evaluation}
\label{ExperimentalResults}

The evaluation focuses on two aspects. The first part determines whether audit events produced by a live K8s cluster are successfully processed by the monitoring pipeline and converted into alerts when they violate the TLA\textsuperscript{+} multitenancy model. We evaluate this using controlled, separate scenarios, where each script targets a specific policy violation. The second part of the experiment involves generating workloads of different sizes (10, 50, 100, 1000) for different number of tenants (2, 5, 10) to showcase the pipeline's scalability.

\subsection{Environment Configuration and Methodology}
The experiment was conducted on a local \texttt{kubeadm} K8s cluster running on three \texttt{libvirt} virtual machines: one control-plane node and two worker nodes. The deployment of the monitoring tool includes all parts described in the previous section: the webhook receiver, the event filtering component, the NATS messaging layer, and the TLC-based trace checker.

\subsubsection*{Validation Experiments}

We evaluated 5 different scenarios that corresponds to the actions in the multitenancy TLA\textsuperscript{+} model. For each scenario, the expected result is one or more alerts in the NATS \texttt{Alerts} stream. Each alert references the audit ID and the type of event that triggered it. When the expected alert is produced, the experiment validates the complete monitoring path: the API server emits an audit event, the webhook receives it, the event is filtered and converted, the trace checker processes it, and the violation is reported. Before each scenario, the required tenant resources are created from scratch. After each scenario finishes, the created resources are removed.

The results in the repository show that the monitoring pipeline was able to detect each of the evaluated RBAC-related multitenancy violations.
\vspace{0.2cm}

{
\begin{minipage}{0.47\textwidth}
\begin{lstlisting}
apiVersion: rbac.authorization.k8s.io/v1
kind: ClusterRole
metadata:
  name: dev
rules:
- apiGroups: [""]
  resources: ["pods", "configmaps", 
    "secrets"]
  verbs: ["get", "list", "create", 
    "update", "patch", "delete"]


\end{lstlisting}
\vspace{1.14cm}
\end{minipage}
\hfill
\begin{minipage}{0.47\textwidth}
\begin{lstlisting}
apiVersion: rbac.authorization.k8s.io/v1
kind: RoleBinding
metadata:
  name: tenant-b-binding
  namespace: tenant-b
subjects:
- kind: Group
  name: tenant-a  # should be tenant-b
  apiGroup: rbac.authorization.k8s.io
roleRef:
  kind: ClusterRole
  name: dev
  apiGroup: rbac.authorization.k8s.io
\end{lstlisting}
\end{minipage}
    \captionof{lstlisting}{Simplified cross-tenant access violation example.}
    \label{lst:cross-tenant-example}
}

\vspace{0.2cm}
Listing~\ref{lst:cross-tenant-example} showcases the K8s events created by the \texttt{01-cross-tenant-access.sh} script. A ClusterRole, \texttt{dev}, grants access to Pods, ConfigMaps and Secrets. The \texttt{RoleBinding} in \texttt{Namespace} \texttt{tenant-b} wrongly binds the \texttt{tenant-a} group instead of \texttt{tenant-b}.

As a result of the misconfiguartion, \texttt{tenant-a-user}'s access request is incorrectly authorized and they can successfully access resources in \texttt{tenant-b}. The monitoring pipeline detects the resulting cross-tenant access and emits the following alert: \lstinline[language=Java]
|[["54e2c017-14f6-4ee5-a4bf-202e9d249362","access.attempt"]]|.

The first element of the alert is the K8s \texttt{auditID}, which uniquely identifies the corresponding audit event. An administrator can then retrieve the complete audit record for further investigation. An excerpt of the matching audit event is shown in Listing~\ref{lst:audit-example}.



\begin{lstlisting}[float=h, 
language=Java,
label={lst:audit-example},
caption={Audit log extract}
]

{
  "auditID": "54e2c017-14f6-4ee5-a4bf-202e9d249362",
  "verb": "list",
  "user": {
    "username": "tenant-a-user",
    "groups": ["tenant-a"]
  },
  "objectRef": {
    "resource": "pods",
    "namespace": "tenant-b"
  },
  "responseStatus": {
    "code": 200
  },
  "annotations": {
    "authorization.k8s.io/reason": "RBAC: allowed by RoleBinding \"tenant-b-binding/tenant-b\" of ClusterRole \"dev\" to Group \"tenant-a\""
  }
}

\end{lstlisting}

\subsubsection*{Scalability Experiment}

\begin{figure}[h]
\centering
\resizebox{0.98\textwidth}{!}{%
\begin{tikzpicture}[
    node distance=0.7cm and 1.0cm,
    block/.style={
        draw,
        rounded corners,
        align=center,
        minimum width=3cm,
        minimum height=3.7cm,
        font=\huge\sffamily,
        fill=white,
        inner sep=4pt
    },
    group/.style={
        draw,
        dashed,
        inner sep=0.35cm,
        rounded corners,
        fill=blue!5,
        fill opacity=0.15
    },
    timing/.style={
        font=\huge\sffamily\itshape,
        align=center
    },
    ->, >=stealth, thick
]

\node[block] (prepare) {
Prepare\\
isolated run
};

\node[block, right=of prepare] (start) {
Start experiment timer\\
and per-batch TLC timing
};

\node[block, right=of start] (generate) {
Generate $N$\\
workload actions
};

\node[block, right=of generate] (drain) {
Wait for AUDIT and\\
AUDIT\_MT consumer drain
};

\node[block, right=of drain] (tlcwait) {
Wait for active\\
TLC batch completion
};

\node[block, right=of tlcwait] (stop) {
Stop timing collection\\
and experiment timer
};

\draw (prepare) -- (start);
\draw (start) -- (generate);
\draw (generate) -- (drain);
\draw (drain) -- (tlcwait);
\draw (tlcwait) -- (stop);

\draw[<->, dashed]
    ([yshift=0.65cm]generate.north west)
    --
    node[midway, above, timing] {
        Input generation duration
    }
    ([yshift=0.65cm]generate.north east);

\draw[<->, dashed]
    ([yshift=-0.85cm]start.south west)
    --
    node[midway, below, timing] {
        Total experiment duration
    }
    ([yshift=-0.85cm]stop.south east);

\begin{scope}[on background layer]
    \node[
        group,
        fit=(prepare) (start) (generate) (drain) (tlcwait) (stop)
    ] (experiment) {};
\end{scope}

\end{tikzpicture}%
}
\vspace{-0.5em}
\caption{
Execution flow for an experiment run.
}
\label{fig:scalability_experiment_methodology}
\end{figure}
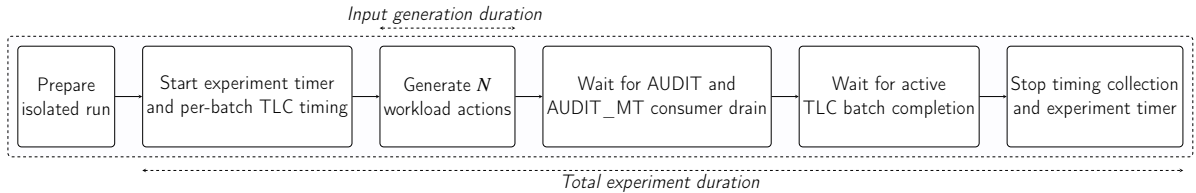

For this experiment, the goal is to measure the scalability of the monitoring pipeline. Figure~\ref{fig:scalability_experiment_methodology} describes the steps followed during each run. Each run is isolated, and multiple timers are used to measure both the overall experiment duration and the processing time of individual TLC batches.

We measured the following metrics:
\begin{compactitem}
\item \emph{Input generation time} ($T_{\mathrm{gen}}$) represents the wall-clock time required to generate the configured number of workload actions. 

\item \emph{Experiment total time} ($T_{\mathrm{exp}}$) is the complete end-to-end duration.

\item \emph{TLC duration} ($T_{\mathrm{TLC}}$) is the accumulated duration of the TLC invocations.

\item \emph{TLC time spent on messages}, the ($T_{\mathrm{nonfetch}}$) divided by no. of messages:
\[
        T_{\mathrm{nonfetch/msg}} = 
        \frac{T_{\mathrm{TLC,non\mbox{-}fetch}}}
         {msgs}.
\]

\item \emph{TLC non-fetch time} ($T_{\mathrm{nonfetch}}$) is the TLC invocation duration minus measured NATS fetch time.

\item \emph{TLC real-time factor ratio} between TLC non-fetch time and input-generation time,
\[
    R_{\mathrm{TLC}} =
    \frac{T_{\mathrm{TLC,non\mbox{-}fetch}}}
         {T_{\mathrm{generation}}}.
\]

Values below $1$ indicate that the measured TLC processing cost is shorter than the workload-generation interval, while values above $1$ indicate that TLC requires more processing time than the workload takes to generate.

\item \emph{End-to-end factor} is the ratio between total experiment duration and input-generation time,
\[
    R_{\mathrm{E2E}} =
    \frac{T_{\mathrm{experiment}}}
         {T_{\mathrm{generation}}}.
\]

This ratio suggests whether the pipeline could fall behind the workload generation in time.

\item \emph{TLC messages and batches} -- the mean number of audit messages processed by TLC and the mean number of TLC batches required. 
The workload size is the number of Kubernetes activities generated during an experiment. The amount of messages eventually handled by TLC may range from this value and vary among runs and tenant configurations, as only audit events relevant to the multitenancy architecture are transmitted for verification.

\item \emph{Standard deviation} is the run-to-run variability around the reported mean for each configuration.

\end{compactitem}

In order to adapt its processing behavior to the monitored cluster, the monitoring pipeline provides a number of adjustable options. In the evaluated configuration, TLC requests batches of up to 50 messages, with a pull-request expiration time of 4.5 seconds. The consumer waits 4.5 seconds for a batch to be filled, or it is processed right away if 50 messages become available before the expiry period; if not, the messages gathered during that time are processed as a smaller batch. These settings  may be modified based on the event rate and latency needs of a specific cluster.

\subsection{Results}
Tables~\ref{tab:scalability_tlc_all} and ~\ref{tab:scalability_tlc_all_2} describe the results of the scalability experiment. Both tables include the number of tenants, actions, and runs. The first table focuses on TLC processing characteristics: the number of processed messages, batching behavior, total processing time, non-fetch time, and non-fetch processing time per message. The second table focuses on workload generation and the overall timing of the experiment. It presents $T_{\mathrm{gen}}$, $T_{\mathrm{exp}}$, and the ratios used to assess the real-time behavior of the monitoring pipeline.

\begin{table*}[h!]
\centering
\scriptsize
\setlength{\tabcolsep}{3pt}
\resizebox{\textwidth}{!}{%
\begin{tabular}{rrrrrrrrr}
\hline
Tenants &
Actions &
Runs &
Msgs. &
Batches &
Msgs./batch &
$T_{\mathrm{TLC}}$ (ms) &
$T_{\mathrm{nonfetch}}$ (ms) &
$T_{\mathrm{nonfetch/msg}}$ (ms) \\
\hline

2 & 10   & 30 & 15.77   & 1.27  & 12.45 &
$7076.0 \pm 2495.5$ &
$1290.3 \pm 445.3$ &
$89.8 \pm 58.2$ \\

2 & 50   & 20 & 56.10   & 2.05  & 27.37 &
$9738.3 \pm 3153.4$ &
$2180.2 \pm 622.3$ &
$38.7 \pm 9.8$ \\

2 & 100  & 10 & 107.10  & 3.00  & 35.70 &
$13574.6 \pm 2416.1$ &
$3280.1 \pm 699.4$ &
$30.4 \pm 4.5$ \\

2 & 1000 & 10 & 1071.20 & 22.30 & 48.04 &
$71149.9 \pm 5019.3$ &
$26206.3 \pm 1062.2$ &
$24.5 \pm 0.8$ \\

\hline

5 & 10   & 30 & 13.57  & 1.30  & 10.44 &
$7330.9 \pm 2574.2$ &
$1386.1 \pm 456.5$ &
$103.2 \pm 26.5$ \\

5 & 50   & 20 & 41.50  & 1.15  & 36.09 &
$6495.7 \pm 2140.3$ &
$1320.4 \pm 384.8$ &
$32.3 \pm 9.5$ \\

5 & 100  & 10 & 82.50  & 2.40  & 34.38 &
$13218.9 \pm 2896.7$ &
$2730.5 \pm 500.4$ &
$33.1 \pm 5.5$ \\

5 & 1000 & 10 & 775.90 & 16.50 & 47.02 &
$66235.1 \pm 2674.3$ &
$20069.7 \pm 684.8$ &
$25.9 \pm 0.8$ \\

\hline

10 & 10   & 30 & 12.00  & 1.00  & 12.00 &
$7292.3 \pm 166.6$ &
$2724.2 \pm 166.8$ &
$233.4 \pm 42.1$ \\

10 & 50   & 20 & 36.60  & 1.05  & 34.86 &
$7795.6 \pm 1167.5$ &
$3080.8 \pm 638.2$ &
$85.5 \pm 15.8$ \\

10 & 100  & 10 & 68.70  & 2.00  & 34.35 &
$14229.4 \pm 594.3$ &
$6122.2 \pm 511.8$ &
$89.3 \pm 5.8$ \\

10 & 1000 & 10 & 689.60 & 14.30 & 48.22 &
$71025.4 \pm 1377.4$ &
$48484.2 \pm 1430.4$ &
$70.3 \pm 1.7$ \\

\hline

\end{tabular}%
}
\caption{TLC processing characteristics across tenant configurations.}
\label{tab:scalability_tlc_all}
\end{table*}

\begin{table*}[h!]
\centering
\scriptsize
\setlength{\tabcolsep}{4pt}
\resizebox{\textwidth}{!}{%
\begin{tabular}{rrrrrrr}
\hline
Tenants &
Actions &
Runs &
$T_{\mathrm{gen}}$ (ms) &
$R_{\mathrm{TLC}}$ &
$R_{\mathrm{E2E}}$ &
$T_{\mathrm{exp}}$ (ms) \\
\hline

2 & 10   & 30 &
$1046.5 \pm 188.6$ &
$1.263 \pm 0.473$ &
$7.375 \pm 1.516$ &
$7579.1 \pm 1251.4$ \\

2 & 50   & 20 &
$4364.0 \pm 1129.4$ &
$0.520 \pm 0.172$ &
$2.609 \pm 0.645$ &
$11017.2 \pm 2338.0$ \\

2 & 100  & 10 &
$8998.2 \pm 1159.0$ &
$0.366 \pm 0.077$ &
$1.763 \pm 0.186$ &
$15744.8 \pm 1621.1$ \\

2 & 1000 & 10 &
$86005.1 \pm 4406.9$ &
$0.305 \pm 0.020$ &
$1.073 \pm 0.018$ &
$92302.5 \pm 4752.9$ \\

\hline

5 & 10   & 30 &
$1627.3 \pm 213.9$ &
$0.867 \pm 0.310$ &
$4.751 \pm 0.677$ &
$7636.1 \pm 728.4$ \\

5 & 50   & 20 &
$4345.2 \pm 315.2$ &
$0.305 \pm 0.093$ &
$1.788 \pm 0.180$ &
$7724.9 \pm 468.4$ \\

5 & 100  & 10 &
$8758.8 \pm 906.9$ &
$0.312 \pm 0.045$ &
$1.621 \pm 0.159$ &
$14085.9 \pm 747.3$ \\

5 & 1000 & 10 &
$75264.7 \pm 2627.3$ &
$0.267 \pm 0.007$ &
$1.100 \pm 0.013$ &
$82783.7 \pm 3301.1$ \\

\hline

10 & 10   & 30 &
$2747.5 \pm 162.0$ &
$0.994 \pm 0.076$ &
$4.134 \pm 0.281$ &
$11364.0 \pm 1032.9$ \\

10 & 50   & 20 &
$5755.0 \pm 709.9$ &
$0.541 \pm 0.119$ &
$2.110 \pm 0.286$ &
$12024.9 \pm 1370.1$ \\

10 & 100  & 10 &
$9454.0 \pm 800.9$ &
$0.651 \pm 0.069$ &
$2.064 \pm 0.136$ &
$19420.0 \pm 576.6$ \\

10 & 1000 & 10 &
$78971.8 \pm 895.5$ &
$0.614 \pm 0.017$ &
$1.124 \pm 0.019$ &
$88718.8 \pm 1433.3$ \\

\hline
\end{tabular}%
}
\caption{End-to-end scalability and real-time processing characteristics across tenant configurations.}
\label{tab:scalability_tlc_all_2}
\end{table*}

\subsection{Discussion}
The end-to-end factor ($R_{\mathrm{E2E}}$) gets lower as the action count increases, which suggests that the overhead of the pipeline becomes less significant under larger workloads. For smaller batches, the 4.5-second pull expiration is highly influential because batches are often only partially filled. In larger workloads, batches become fuller and the processing cost per message generally decreases. Similarly, the TLC real-time factor ($R_{\mathrm{TLC}}$) lowers as the workload becomes more increased.

The $T_{\mathrm{nonfetch/msg}}$ entries suggest that processing an individual audit event becomes more expensive as the modeled cluster state grows. We can observe this in the results for the ten-tenant configuration. A possible explanation is that larger modeled domains and state structures increase the cost of TLC's internal state processing.

\section{Related Work}
\label{RelatedWork}
This section presents related work in the fields of K8s multitenancy and trace checking. For the former, we present open-souce tools that propose different solutions to the problem. For the latter, we present peer-reviewed papers and one open-source project.

\subsection{K8s Multitenancy}

\emph{Kyverno} is a K8s-native policy engine that is used as an admission controller (see \cite{KyvernoDocs}). It connects to the K8s API Server using an admission webhook.  \emph{OPA Gatekeeper} (see \cite{GatekeeperDocs}) is another K8s policy engine used to enforce policies at admission-time. It is similar to Kyverno in the sense that both tools can evaluate requests sent to the K8s API server before resources are persisted. Kyverno and OPA Gatekeeper can prevent or report resources that violate K8s policies, but they do not provide a formal model of the expected multitenant state or reason over audit-log traces in the same way as our proposed tool. The monitoring pipeline is not a replacement for admission controllers: they evaluate resources against policies, while the proposed tool checks audit-log events against a formal multitenancy model.

\emph{Falco} is a cloud native security tool focused on threat detection (see \cite{FalcoDocs}). Unlike Kyverno and OPA Gatekeeper, which are mainly used to evaluate K8s resources at admission time, Falco observes runtime events and raises alerts in case it detects suspicious behavior. In multitenant environments, Falco can complement preventive isolation mechanisms by detecting suspicious cross-tenant behavior. Both the pipeline and Falco consume event streams and produce alerts, but the goals are different: Falco detects suspicious behavior using predefined or custom security rules, while the proposed pipeline checks K8s audit events against a formal multitenancy model.

\emph{Capsule} is a K8s multitenancy framework that introduces a higher level Tenant abstraction (for more details: \cite{CapsuleDocs}). A Tenant is a cluster-scoped resource that groups one or more Namespaces under the same administrative boundary. Rather than validating resources or detecting suspicious runtime actions, Capsule changes how namespace-based multitenancy is represented in a cluster. Capsule has a different scope from the monitoring pipeline: our tool does not introduce a new abstraction in K8s, rather it utilizes existing cluster information to detect multitenancy violations.

\subsection{Trace Checking}

Howard et al.~\cite{Howard2011} present a case study that suggests the feasibility of replaying execution traces against formal specifications and highlights challenges related to trace processing. Their approach is primarily offline, as traces are gathered and analyzed after execution. On the other hand, our system is designed for continuous cluster monitoring rather than a retrospective analysis tool.

Kuppe et al.\cite{ValidateTraces} present a trace validation framework that links distributed Java program executions to high-level TLA\textsuperscript{+} specifications through code instrumentation. The authors create a trace-specific TLA\textsuperscript{+} wrapper around the base protocol spec and use TLC to measure the impact of trace detail on TLC's state-space search cost. Our tool allows TLA\textsuperscript{+}-aided monitoring to be integrated directly into the operational lifecycle of an active cluster. However, we do not allow incomplete traces: every state must be explicitly mentioned in the cluster logs. This is not generally a problem for K8s audit logs, since they represent a history of the cluster state.

Howard et al. \cite{howard2024smartcasualverificationconfidential} show how TLA\textsuperscript{+} can be applied to large-scale production systems using bounded model checking, simulation, and offline trace validation. Their work validates a framework by matching instrumented execution traces against low-level consensus and high-level consistency specifications. The validation is mainly performed offline in controlled test environments, rather than used as an online runtime monitor.

Rahman et al. \cite{ExtremeModeling2020} present an industrial case study of applying TLA\textsuperscript{+} and TLC for model-based trace checking in MongoDB. Their study highlights challenges in tracing highly concurrent systems: hierarchical locking, state visibility, and the difficulty of obtaining consistent snapshots of internal state without altering system behavior. In contrast to this approach, our work relies exclusively on K8s audit logs, which provide a complete and externally visible record of control-plane actions, and does not require changes in application code.

Ding et al. \cite{Ding} propose a runtime verification tool written in Go called Ellsberg, which emits alerts when a distributed protocol's implementation outputs messages inconsistent with a protocol specification. The workflow starts from a TLA\textsuperscript{+} specification, from which users should derive a specification that Ellsberg can use. In their work, they modified the systems' network APIs to forward protocol messages to the tool. In contrast, our pipeline directly uses TLC and TLA\textsuperscript{+} specifications to check observed behavior inferred from a cluster's audit logs.

The Open Network Operating System (ONOS) TLA\textsuperscript{+} Monitor~\cite{ONOSMonitor} is a public conformance monitoring tool associated with the \(\mu\)ONOS platform. Unlike most related work presented in this section, it is not presented as a peer-reviewed paper, but through public documentation and source code. The official ONOS documentation describes conformance monitoring as a mechanism for checking whether \(\mu\)ONOS services abide by formal specifications in near real time. For more details, see \cite{ONOSDocs}. The ONOS monitor showcases a practical use of TLA\textsuperscript{+} outside offline model checking, since formal specifications are used to validate executions of a running system. The ONOS monitor focuses on a SDN platform, while our pipeline targets K8s clusters. For the trace processing, the ONOS monitor uses uses a sliding window batching strategy. In our implementation we use explicit batching and checkpointing over the audit-event stream.

\section{Conclusions and Future Work}
\label{ConclusionFW}

In this paper, we have shown that a formal state-machine-based specification can be used as the basis for runtime monitoring in a K8s cluster. The main idea was to link a TLA\textsuperscript{+} multitenancy model to events produced by a live cluster in order to check observed system activity against the rules expressed in the specification. We packaged the pipeline as a K8s-native deployment, configured a K8s cluster, deployed the pipeline, and evaluated it through a set of experimental scenarios. The experiment showed that the selected multitenancy violations can be detected from audit events in a live cluster and verified the whole path: from API server audit logging, to webhook ingestion, filtering, trace checking, and alert generation. Overall, this work demonstrates that TLA\textsuperscript{+} can be used not only as a design-time specification language, but also as the basis for a monitoring pipeline connected to real system traces. Using the same language for both modeling the desired state and catching events that violate this state increases the usefulness of the specifications, especially in cases where the implementation can easily drift from the intended design.

However, our work also presents limitations: it depends on access to K8s audit logging and audit webhook configuration. This may not be an issue in self-managed clusters where administrators can modify the API server configuration, but it is not directly applicable to managed K8s environments where audit webhook configuration is not exposed to the users. Moreover, the multitenancy specification is not universally reusable since it requires individual adaptation.

For future work, we plan to extend the model to cover additional multitenancy-related concerns, such as resource consumption and \texttt{NetworkPolicies}. Supporting these aspects would also require the pipeline to ingest additional sources of runtime information, such as pod- and container-level logs.

\bibliographystyle{eptcsalpha}
\bibliography{bibliography}

\newcommand{\etalchar}[1]{$^{#1}$}
\begin{thebibliography}{{ONO}26b}
\providecommand{\bibitemdeclare}[2]{}
\providecommand{\surnamestart}{}
\providecommand{\surnameend}{}
\providecommand{\urlprefix}{Available at }
\providecommand{\url}[1]{\texttt{#1}}
\providecommand{\href}[2]{\texttt{#2}}
\providecommand{\urlalt}[2]{\href{#1}{#2}}
\providecommand{\doi}[1]{doi:\urlalt{https://doi.org/#1}{#1}}
\providecommand{\eprint}[1]{arXiv:\urlalt{https://arxiv.org/abs/#1}{#1}}
\providecommand{\bibinfo}[2]{#2}

\bibitemdeclare{inproceedings}{batsonleslie2002}
\bibitem[BL02]{batsonleslie2002}
\bibinfo{author}{Brannon \surnamestart Batson\surnameend} \&
  \bibinfo{author}{Leslie \surnamestart Lamport\surnameend}
  (\bibinfo{year}{2002}): \emph{\bibinfo{title}{High-Level Specifications:
  Lessons from Industry}}.
\newblock \bibinfo{volume}{2852}, pp. \bibinfo{pages}{242--261},
  \doi{10.1007/978-3-540-39656-7\_10}.

\bibitemdeclare{inproceedings}{ValidateTraces}
\bibitem[CKLM25]{ValidateTraces}
\bibinfo{author}{Horatiu \surnamestart Cirstea\surnameend},
  \bibinfo{author}{Markus~A. \surnamestart Kuppe\surnameend},
  \bibinfo{author}{Benjamin \surnamestart Loillier\surnameend} \&
  \bibinfo{author}{Stephan \surnamestart Merz\surnameend}
  (\bibinfo{year}{2025}): \emph{\bibinfo{title}{Validating Traces of
  Distributed Programs Against TLA+ Specifications}}.
\newblock In \bibinfo{editor}{Alexandre \surnamestart Madeira\surnameend} \&
  \bibinfo{editor}{Alexander \surnamestart Knapp\surnameend}, editors:
  {\slshape \bibinfo{booktitle}{Software Engineering and Formal Methods}},
  \bibinfo{publisher}{Springer Nature Switzerland}, \bibinfo{address}{Cham}, p.
  \bibinfo{pages}{126–143}, \doi{10.1007/978-3-031-77382-2\_8}.
\newblock
  \urlprefix\url{https://link.springer.com/chapter/10.1007/978-3-031-77382-2\_8}.

\bibitemdeclare{article}{ExtremeModeling2020}
\bibitem[DHS20]{ExtremeModeling2020}
\bibinfo{author}{A.~Jesse~Jiryu \surnamestart Davis\surnameend},
  \bibinfo{author}{Max \surnamestart Hirschhorn\surnameend} \&
  \bibinfo{author}{Judah \surnamestart Schvimer\surnameend}
  (\bibinfo{year}{2020}): \emph{\bibinfo{title}{Extreme modelling in
  practice}}.
\newblock {\slshape \bibinfo{journal}{Proc. VLDB Endow.}}
  \bibinfo{volume}{13}(\bibinfo{number}{9}), p. \bibinfo{pages}{1346–1358},
  \doi{10.14778/3397230.3397233}.
\newblock \urlprefix\url{http://vldb.org/pvldb/vol13/p1346-davis.pdf}.

\bibitemdeclare{inproceedings}{Ding}
\bibitem[DWLP25]{Ding}
\bibinfo{author}{Ding \surnamestart Ding\surnameend}, \bibinfo{author}{Zhanghan
  \surnamestart Wang\surnameend}, \bibinfo{author}{Jinyang \surnamestart
  Li\surnameend} \& \bibinfo{author}{Aurojit \surnamestart Panda\surnameend}
  (\bibinfo{year}{2025}): \emph{\bibinfo{title}{{Runtime Protocol Refinement
  Checking for Distributed Protocol Implementations}}}.
\newblock In: {\slshape \bibinfo{booktitle}{22nd USENIX Symposium on Networked
  Systems Design and Implementation (NSDI 25)}}, \bibinfo{publisher}{USENIX
  Association}, \bibinfo{address}{Philadelphia, PA}, pp.
  \bibinfo{pages}{1305--1326}, \doi{10.5555/3767955.3768025}.
\newblock
  \urlprefix\url{https://www.usenix.org/conference/nsdi25/presentation/ding}.

\bibitemdeclare{misc}{FalcoDocs}
\bibitem[{Fal}26]{FalcoDocs}
\bibinfo{author}{\surnamestart {Falco Authors}\surnameend}
  (\bibinfo{year}{2026}): \emph{\bibinfo{title}{{Falco Documentation}}}.
\newblock \bibinfo{howpublished}{\url{https://falco.org/docs/}}.
\newblock \bibinfo{note}{Accessed: 2026-04-26}.

\bibitemdeclare{misc}{HelmDocs}
\bibitem[{Hel}26]{HelmDocs}
\bibinfo{author}{\surnamestart {Helm Authors}\surnameend}
  (\bibinfo{year}{2026}): \emph{\bibinfo{title}{{Helm Documentation}}}.
\newblock \bibinfo{howpublished}{\url{https://helm.sh/docs/}}.
\newblock \bibinfo{note}{Accessed: 2026-06-15}.

\bibitemdeclare{article}{Howard2011}
\bibitem[HGG{\etalchar{+}}11]{Howard2011}
\bibinfo{author}{Yvonne \surnamestart Howard\surnameend},
  \bibinfo{author}{Stefan \surnamestart Gruner\surnameend},
  \bibinfo{author}{A.~\surnamestart Gravell\surnameend}, \bibinfo{author}{Carla
  \surnamestart Ferreira\surnameend} \& \bibinfo{author}{Juan \surnamestart
  {Augusto Wrede}\surnameend} (\bibinfo{year}{2011}):
  \emph{\bibinfo{title}{{Model-Based Trace-Checking}}}.
\newblock {\slshape \bibinfo{journal}{CoRR}} \bibinfo{volume}{abs/1111.2825},
  \doi{10.48550/arXiv.1111.2825}.
\newblock \urlprefix\url{https://arxiv.org/abs/1111.2825}.

\bibitemdeclare{misc}{howard2024smartcasualverificationconfidential}
\bibitem[HKA{\etalchar{+}}24]{howard2024smartcasualverificationconfidential}
\bibinfo{author}{Heidi \surnamestart Howard\surnameend},
  \bibinfo{author}{Markus~A. \surnamestart Kuppe\surnameend},
  \bibinfo{author}{Edward \surnamestart Ashton\surnameend},
  \bibinfo{author}{Amaury \surnamestart Chamayou\surnameend} \&
  \bibinfo{author}{Natacha \surnamestart Crooks\surnameend}
  (\bibinfo{year}{2024}): \emph{\bibinfo{title}{{Smart Casual Verification of
  the Confidential Consortium Framework}}},
  \doi{https://doi.org/10.48550/arXiv.2406.17455}.
\newblock \eprint{2406.17455}.

\bibitemdeclare{misc}{K8sDocs}
\bibitem[{Kub}26a]{K8sDocs}
\bibinfo{author}{\surnamestart {Kubernetes Authors}\surnameend}
  (\bibinfo{year}{2026}): \emph{\bibinfo{title}{{Kubernetes Documentation}}}.
\newblock \bibinfo{howpublished}{\url{https://kubernetes.io/}}.
\newblock \bibinfo{note}{Accessed: 2026-04-25}.

\bibitemdeclare{misc}{K8sAudit}
\bibitem[{Kub}26b]{K8sAudit}
\bibinfo{author}{\surnamestart {Kubernetes Authors}\surnameend}
  (\bibinfo{year}{2026}): \emph{\bibinfo{title}{{Kubernetes Documentation}:
  Auditing}}.
\newblock
  \bibinfo{howpublished}{\url{https://kubernetes.io/docs/tasks/debug/debug-cluster/audit/}}.
\newblock \bibinfo{note}{Accessed: 2026-04-25}.

\bibitemdeclare{misc}{K8sRBACPractices}
\bibitem[{Kub}26c]{K8sRBACPractices}
\bibinfo{author}{\surnamestart {Kubernetes Authors}\surnameend}
  (\bibinfo{year}{2026}): \emph{\bibinfo{title}{{Kubernetes Documentation: RBAC
  Good Practices}}}.
\newblock
  \bibinfo{howpublished}{\url{https://kubernetes.io/docs/concepts/security/rbac-good-practices/}}.
\newblock \bibinfo{note}{Accessed: 2026-05-27}.

\bibitemdeclare{misc}{KyvernoDocs}
\bibitem[{Kyv}26]{KyvernoDocs}
\bibinfo{author}{\surnamestart {Kyverno Authors}\surnameend}
  (\bibinfo{year}{2026}): \emph{\bibinfo{title}{{Kyverno Documentation}:
  Introduction}}.
\newblock \bibinfo{howpublished}{\url{https://kyverno.io/docs/introduction/}}.
\newblock \bibinfo{note}{Accessed: 2026-04-26}.

\bibitemdeclare{book}{SpecifyingSystems}
\bibitem[Lam02]{SpecifyingSystems}
\bibinfo{author}{Leslie \surnamestart Lamport\surnameend}
  (\bibinfo{year}{2002}): \emph{\bibinfo{title}{{Specifying Systems: The
  TLA$^+$ Language and Tools for Hardware and Software Engineers}}}.
\newblock \bibinfo{publisher}{Addison-Wesley}.
\newblock
  \urlprefix\url{https://lamport.azurewebsites.net/tla/book-02-08-08.pdf}.

\bibitemdeclare{misc}{NatsJetstream}
\bibitem[{NAT}26]{NatsJetstream}
\bibinfo{author}{\surnamestart {NATS Maintainers}\surnameend}
  (\bibinfo{year}{2026}): \emph{\bibinfo{title}{{NATS Documentation}:
  JetStream}}.
\newblock
  \bibinfo{howpublished}{\url{https://docs.nats.io/nats-concepts/jetstream}}.
\newblock \bibinfo{note}{Accessed: 2026-06-13}.

\bibitemdeclare{article}{Newcombe2015}
\bibitem[NRZ{\etalchar{+}}15]{Newcombe2015}
\bibinfo{author}{Chris \surnamestart Newcombe\surnameend}, \bibinfo{author}{Tim
  \surnamestart Rath\surnameend}, \bibinfo{author}{Fan \surnamestart
  Zhang\surnameend}, \bibinfo{author}{Bogdan \surnamestart
  Munteanu\surnameend}, \bibinfo{author}{Marc \surnamestart Brooker\surnameend}
  \& \bibinfo{author}{Michael \surnamestart Deardeuff\surnameend}
  (\bibinfo{year}{2015}): \emph{\bibinfo{title}{{How Amazon Web Services uses
  formal methods}}}.
\newblock {\slshape \bibinfo{journal}{Communications of the ACM}},
  \doi{10.1145/2699417}.
\newblock
  \urlprefix\url{https://www.amazon.science/publications/how-amazon-web-services-uses-formal-methods}.

\bibitemdeclare{misc}{ONOSDocs}
\bibitem[{ONO}26a]{ONOSDocs}
\bibinfo{author}{\surnamestart {ONOS Project}\surnameend}
  (\bibinfo{year}{2026}): \emph{\bibinfo{title}{{ONOS Documentation: Open
  Network Operating System}}}.
\newblock \bibinfo{howpublished}{\url{https://docs.onosproject.org/}}.
\newblock \bibinfo{note}{Accessed: 2026-06-23}.

\bibitemdeclare{misc}{ONOSMonitor}
\bibitem[{ONO}26b]{ONOSMonitor}
\bibinfo{author}{\surnamestart {ONOS Project}\surnameend}
  (\bibinfo{year}{2026}): \emph{\bibinfo{title}{{TLA+ {M}onitor}}}.
\newblock
  \bibinfo{howpublished}{\url{https://github.com/onosproject/tlaplus-monitor}}.
\newblock \bibinfo{note}{Accessed: 2026-04-25}.

\bibitemdeclare{misc}{GatekeeperDocs}
\bibitem[{Ope}26]{GatekeeperDocs}
\bibinfo{author}{\surnamestart {Open Policy Agent Contributors}\surnameend}
  (\bibinfo{year}{2026}): \emph{\bibinfo{title}{{Gatekeeper Documentation}:
  Introduction}}.
\newblock
  \bibinfo{howpublished}{\url{https://open-policy-agent.github.io/gatekeeper/website/docs/}}.
\newblock \bibinfo{note}{Accessed: 2026-04-26}.

\bibitemdeclare{misc}{CapsuleDocs}
\bibitem[{Pro}26]{CapsuleDocs}
\bibinfo{author}{\surnamestart {Project Capsule Authors}\surnameend}
  (\bibinfo{year}{2026}): \emph{\bibinfo{title}{{Capsule Documentation}:
  Overview}}.
\newblock
  \bibinfo{howpublished}{\url{https://projectcapsule.dev/docs/overview/}}.
\newblock \bibinfo{note}{Accessed: 2026-06-13}.

\end{thebibliography}
\end{document}